\documentclass[12pt, hyper]{article}
\usepackage{color}
\usepackage[table]{xcolor}
\usepackage{amsmath}
\usepackage{amsfonts}
\usepackage{amsopn}
\usepackage{graphicx} 
\usepackage{adjustbox}
\usepackage{amssymb}
\usepackage{amsthm}
\usepackage{hyperref}
\usepackage{tikz-cd}
\usepackage{dynkin-diagrams}
\usepackage{mathrsfs}
\DeclareMathAlphabet{\mathpzc}{OT1}{pzc}{m}{it}
\usepackage{verbatim}
\usepackage{multirow}
\usepackage{enumitem}
\usepackage{setspace} 
\usepackage{arydshln}
\usepackage{cite}
\usepackage{amsthm}
\usepackage[margin=0.5cm]{caption}   
\usepackage[font={small}]{caption}
\usepackage{cleveref}
\newcommand{\be}{\begin{equation}}
\newcommand{\ee}{\end{equation}}
\newtheorem*{thm}{Theorem}

\begin{document}
\usetikzlibrary{calc,decorations.markings}
\tikzset{
    partial ellipse/.style args={#1:#2:#3}{
        insert path={+ (#1:#3) arc (#1:#2:#3)}
    }
}

\begin{titlepage}
\vfill
\begin{flushright}
{\tt\normalsize KIAS-P21065}\\

\end{flushright}
\vfill
\begin{center}
{\large\bf Can the energy bound $E\geq 0$ imply supersymmetry?}

\vskip 1cm

Jin-Beom Bae$^\dagger$, Zhihao Duan$^\natural$ 
and Sungjay Lee$^\natural$

\vskip 5mm
$^\dagger${\it Mathematical Institute, University of Oxford \\
$~~ $Andrew Wiles Building, Radcliffe Observatory Quarter \\
$~~ $Woodstock Road, Oxford, OX2 6GG, U.K.}
\vskip 3mm $^\natural${\it Korea Institute for Advanced Study \\
$~~ $85 Hoegiro, Dongdaemun-Gu, Seoul 02455, Korea}
\end{center}
\vfill

\begin{abstract}
\noindent
We utilize the integrality conjecture to show that the torus partition 
function of a fermionic rational conformal theory in the Ramond-Ramond
sector becomes a constant when the bound $h^R \ge \frac{c}{24}$ is satisfied,
where $h^R$ denote the conformal weights of Ramond states and $c$ is the central charge. 
The constant-valued Ramond-Ramond partition function strongly suggests the presence of 
supersymmetry unless a given theory has free fermions. 
The lower bound $h^R \ge \frac{c}{24}$ can then be identified with the unitarity bound of $\mathcal{N}=1$ supersymmetry. We thus propose that, for rational CFTs without free fermions, 
 $(h^R-c/24) \geq 0$ can imply supersymmetry.

\end{abstract}

\vfill
\end{titlepage}

\renewcommand{\thefootnote}{\#\arabic{footnote}}
\setcounter{footnote}{0}


\paragraph{Introduction}

The energy spectrum of a supersymmetric theory is constrained to be non-negative, $E\geq 0$. 
In contrast, does the lower bound on the energy spectrum 
imply the supersymmetry? The answer is no in general. 
To see this, let us consider a non-supersymmetric quantum system with 
a potential that allows a normalizable ground state. By adding to the Lagrangian a suitable constant energy, one can always make the energy spectrum positive while maintaining the non-supersymmetric nature of the system. However, we propose that for fermionic rational conformal field theories (RCFTs) in two dimensions the answer to the above converse question becomes positive. 

We start with the general fermionic CFT in two-dimensional spacetime. To define a fermionic CFT on a manifold $X$, we need to choose a spin structure on 
$X$ specifying a boundary condition of fermions around each non-trivial cycle. 
For example, there are four different spin structures on a torus $\mathbb{T}^2$. 
In the present work, we focus on the Ramond-Ramond boundary 
condition where fermions are periodic around any cycles 
on $\mathbb{T}^2$. The partition function associated with the Ramond-Ramond (RR) 
boundary condition can be expressed as follows 

\begin{align}
Z_{RR}(\tau,\bar{\tau}) = \mbox{Tr}_{\mathcal{H}_{R}}\Big[ (-1)^F q^{h^R-\frac{c}{24}} \bar{q}^{\bar{h}^R-\frac{c}{24}} \Big],
\end{align}
where ${\cal H}_R$ denotes the Hilbert space of a given fermionic CFT
in the Ramond sector and $q=e^{2\pi i\tau}$. 
The dimension $\Delta$ and the spin $s$ of a 
state in ${\cal H}_R$ can be determined by 
its left and right (Ramond) conformal weights $h^R$ and ${\bar h}^R$,
$\Delta=h^R + {\bar h}^R$ and $s= | h^R - {\bar h}^R|$. 
The rational CFTs refer to CFTs whose partition function 
can be written as a finite sum of products of holomorphic and 
anti-holomorphic functions on $(\tau,\bar \tau)$, 
\begin{align}\label{adf}
Z_{RR}(\tau,\bar{\tau}) = \sum_{i,j=0}^{N-1}  f_i(\tau) \mathcal{M}_{ij} {\bar f}_j(\bar{\tau}).
\end{align}
Here each holomorphic function $f_i(\tau)$ represents the RR conformal character 
for a primary state of conformal weight $h_i^R$. Without loss of generality, 
we can assume that $h_0^R < h_1^R < \cdots < h_{N-1}^R$. We also define the leading exponent of each $f_i(\tau)$ in $q$-expansion as $m_i = h_i^R - \frac{c}{24}$ in what follows,
\begin{align}\label{df}
  f_i(\tau) = q^{m_i} \sum_{a=0}^\infty F_i(a) q^a\,.
\end{align}

For a superconformal CFT, the unitarity leads to 
a lower bound on the conformal weight in the Ramond sector,
\begin{align}
\label{unitary bound}
    \Big\{ G_0, G_0 \Big\} \propto \Big( h^R - \frac{c}{24} \Big) \geq 0,
\end{align}
where $G_{0}$ is the superconformal charge and $c$ is the central charge. 
The main goal of the present work is to show that
$Z_{RR}(\tau,\bar{\tau})$ becomes either constant or 
zero when the lower bound \eqref{unitary bound} is obeyed. In other words, we will utilize the modular constraint of $f_i(\tau)$ to propose constant $Z_{RR}(\tau,\bar{\tau})$ when all the exponents of $f_i(\tau)$ are non-negative.    
To have a constant RR partition function, 
the contributions from bosonic and fermionic states 
other than vacuum have to be cancelled. 
One can hardly expect the above cancellation unless 
the given theory has either supersymmetry or 
free fermions. Note that a free fermion subject 
to the RR boundary condition on $\mathbb{T}^2$ 
has a zero mode, which results in vanishing $Z_{RR}$. 
Thus, we propose that a fermionic RCFT satisfying 
the bound \eqref{unitary bound} is supersymmetric unless 
it has free fermions. In this case, one can regard the constant $Z_{RR}$ 
as the index of supersymmetric RCFTs. 

To support the proposal, we make use of two essential properties 
of the RR characters $f_{i}(\tau)$ of \eqref{adf}. First, the conformal characters
should transform under $\text{SL}(2,\mathbb{Z})$ as the vector-valued 
modular functions; otherwise, the RR partition function cannot be 
modular invariant. Second, each conformal character has 
a integral $q$-expansion whose coefficients count state degeneracies  
weighted by $(-1)^F$. In \cite{Mukhi:2020gnj}, such characters are referred to as quasi-characters. 
In what follows, we provide a proof relying on those properties 
of our claim that $Z_{RR}$ becomes either constant or $0$ when \eqref{unitary bound} is 
satisfied for fermionic RCFTs.

\paragraph{Integrality Conjecture}

It has been known that any vector-valued modular form with $N$-components $f_i(\tau)$ can be understood as independent solutions of a modular invariant linear differential equation (MLDE) of order $N$  \cite{Mathur:1988na}. 
The valence formula applied to the Wronskian of $f_i(\tau)$ then reads 
\begin{align}\label{eq:valence}
  \sum_{i=0}^{N-1} m_i + \frac{l}{6} = \frac{N(N-1)}{12},
\end{align}
where $l$ is the Wronskian index, a non-negative integer except $1$ capturing the number of zeroes of the Wronskian.  
The main idea of classifying two-dimensional RCFT is to pick up the solutions of MLDE that satisfy constraints such as having the positive integer Fourier coefficients, uniqueness of the vacuum, and consistent fusion rule algebra. The classification of the bosonic and fermionic RCFTs has been performed in \cite{Mathur:1988na, Hampapura:2015cea, Hampapura:2016mmz, Chandra:2018pjq,  Mukhi:2019xjy, Kaidi:2021ent, Das:2021uvd, Bae:2020xzl, Bae:2021mej}. In particular, a recent paper \cite{Kaidi:2021ent} utilized the integrality conjecture\cite{atkin1971modular,bantay2007vectorvalued} to provide a new approach to the classification program. 

The modular invariance of $Z_{RR}(\tau,\bar{\tau})$ implies that the RR conformal characters are regarded as a vector-valued modular form. In contrast to the conformal characters of the bosonic RCFTs, the RR conformal characters are allowed to have the negative integer coefficients in $q$-expansion. Nonetheless, the integrality of the coefficients enable us to apply the \textit{integrality conjecture}. According to the integrality conjecture which is proved in \cite{Ng:2012ty} for RCFT characters and recently proved in the general case by \cite{calegari2021unbounded}, 
each RR conformal character should be invariant under the principal congruence subgroup of $\text{SL}(2,\mathbb{Z})$. Such a subgroup is defined with a positive integer $n$ as follows,
\be
\Gamma(n) := \Bigg\{\begin{pmatrix}
a&b\\
s&d
\end{pmatrix} \in \text{SL}(2,\mathbb{Z}),\ \  \begin{pmatrix}
a&b\\
s&d
\end{pmatrix} \equiv \begin{pmatrix}
1&0\\
0&1
\end{pmatrix} \text{mod}\ n\Bigg\}.
\ee
Since we only require the coefficients to be integral, an immediate lesson we learn is that the RR characters also form a representation of the finite group $\text{SL}(2,\mathbb{Z})/\Gamma(n) = \text{SL}(2,\mathbb{Z}_n)$ for an integer $n$. Also we need to remember that the minus of identity matrix in $\text{SL}(2,\mathbb{Z}_n)$ acts trivially on the characters by construction. In other words, we should consider the representation theory of $\text{PSL}(2,\mathbb{Z}_n)$. For small number of characters, 
a plausible strategy then is to classify all possible $n$ for given $N$ and extract information from its representation theory.

Let us state without proof the simplest non-trivial example, i.e., $N = 2$. The finite list of $n$ containing the desired two-dimensional irreducible representation turns out to be 
\be
n \in \{2,6,8,12,20,24,60\}\,.
\ee
A priori, some $n$ in the list does not need to be realized by a rational CFT with two RR characters, but we stress that the converse must be correct. Furthermore, the detail of those representations can be easily accessed from computer software such as GAP \cite{GAP4}. In particular, we can extract from the character table of $\text{SL}(2,\mathbb{Z}_n)$ the exponents of characters $m_i$ as summarized in table \ref{tab_2dexpo}.
For the detailed discussion, we refer the readers to \cite{Kaidi:2021ent}.

\begin{table}[!tp]
\def\arraystretch{1.5}
\begin{tabular}{c|c}
$n$ & Exponents mod 1
\\ \hline
2 & $\left\{0 , {1\over 2} \right\}$\\
6 & $\left\{{2 \over 3} , {1\over 6} \right\}$, $\left\{{1 \over 3} , {5\over 6} \right\}$ \\
8 & $\left\{{1 \over 8} , {3\over 8} \right\},\left\{{5 \over 8} , {7\over 8} \right\}$ \\
12& $\left\{{1 \over 4} , {11\over 12} \right\}, \left\{{3 \over 4} , {5\over 12} \right\}$ $\left\{{1 \over 4} , {7\over 12} \right\}, \left\{{3 \over 4} , {1\over 12} \right\}$ $\left\{{1 \over 12} , {5\over 12} \right\}, \left\{{7 \over 12} , {11\over 12} \right\}$ \\
20& $\left\{{1 \over 20} , {9\over 20} \right\}, \left\{{3 \over 20} , {7\over 20} \right\},\left\{{11 \over 20} , {19\over 20} \right\}, \left\{{13 \over 20} , {17\over 20} \right\}$\\
24& $\left\{{11 \over 24} , {17\over 24} \right\},\left\{{5 \over 24} , {23\over 24} \right\}$ $\left\{{1 \over 24} , {19\over 24} \right\},\left\{{7 \over 24} , {13\over 24} \right\}$ \\
60& $\left\{{11 \over 60} , {59\over 60} \right\},\left\{{17 \over 60} , {53\over 60} \right\}, \left\{{23 \over 60} , {47\over 60} \right\}, \left\{{29 \over 60} , {41\over 60} \right\}$ $\left\{{1 \over 60} , {49\over 60} \right\}, \left\{{7 \over 60} , {43\over 60} \right\},\left\{{19 \over 60} , {31\over 60} \right\}, \left\{{13 \over 60} , {37\over 60} \right\}$
\end{tabular}
\caption{Possible exponents mod 1 for potential RCFTs with two RR characters.}\label{tab_2dexpo}
\end{table}

Now let us come back to our original problem. Combining the valence formula \eqref{eq:valence} when $N = 2$ with the exponents in table \ref{tab_2dexpo}, we easily learn that if RR characters with non-negative exponents are non-trivial, we must have a contradiction. This implies that actually all the characters should vanish, thereby establishing the desired result $Z_{RR}=0$ for the special case of two RR characters. Moreover, all possible exponents mod $1$ up to $N = 5$ are worked out in \cite{Kaidi:2021ent}, which cannot satisfy the corresponding valence formula if we assume they are all non-negative. This gives us an affirmative answer to our question.

While being very concrete, this method would not be practical for general $N$ and we need to look for other approaches. In \cite{Kaidi:2020ecu}, the authors prove a theorem crucial for our purpose. For reader's convenience, we restate it below.
\begin{thm}
Consider an $N$-dimensional weakly holomorphic and integral vector-valued modular function $\vec{f}(q)$ with components $f_i(q)$. Define
\be
\textbf{m} := \text{min} \left(m_0, \cdots, m_{N-1}\right)
\ee
where $f_i(q) \sim q^{m_i}$ near the cusp $i\infty$. If there exists one component 
$f_j(q)$ which is a modular function of $\Gamma(n)$ for some integer $n$ with $m_j \neq 0$, then $\textbf{m} < 0$.
\end{thm}
The proof is not sophisticated, so we briefly outline it here. 
As a modular function for $\Gamma(n)$, the component $f_j(q)$ should obey the valence formula,
\be\label{eq:valence}
\sum_{\tau \in \mathbb{H}/\Gamma(n)} \text{Ord}_\tau (f_j) = 0,
\ee
where $\mathbb{H}$ is the upper half-plane and 
$\text{Ord}_\tau (f_j)$ denotes the leading order of $f_j$ when Laurent expanded around the point $\tau$. 
If $m_j < 0$, the theorem is already proved, so we assume $m_j > 0$. 
Since the order of $f_j(q)$ at $\tau=i\infty$ is $n m_j>0$, the valence formula \eqref{eq:valence}
says that $f_j(q)$ should diverge at some other cusps, denoted by $\tau_\ast$, since a physical conformal character is required to be holomorphic inside $\mathbb{H}/\Gamma(n)$. 
It implies that the $\text{SL}(2,\mathbb{Z})$ invariant partition function \eqref{adf},
\be
Z_{RR}(\tau,\bar{\tau}) = \sum f_i(\tau) \mathcal{M}_{ij} \bar{f}_j(\bar{\tau})\,.
\ee
also diverges at $\tau=\tau_\ast$. Invoking an $\text{SL}(2,\mathbb{Z})$ transformation that maps $\tau_\ast$ to $i\infty$, we readily see that $Z(\tau,\bar{\tau})$ and hence some other component of $\vec f(q)$ diverge at $\tau=i\infty$. In other words, $\textbf{m} < 0 $. This 
is the end of the proof.

In our present situation, recall that we have the constraint $h^R_i \geq \frac{c}{24}$. 
If a given theory has the Ramond spectrum satisfying the bound \eqref{unitary bound}
strictly, i.e., $h^R>c/24$,  
all conformal characters must be trivial and thus the corresponding RR partition function 
is simply zero according to the above theorem. On the other hand, 
let us suppose that the bound is saturated, namely $h^R_\alpha = \frac{c}{24}$. 
The corresponding conformal character can be written as
\be
f_\alpha (q)  = c_\alpha + \tilde{f}_\alpha(q)\,,
\ee 
with $\tilde{f}_\alpha(q)$ now having strictly positive exponent in the $q$-expansion. It is clear that $\tilde{f}_\alpha$ is also modular invariant under the congruence subgroup $\Gamma(n)$ indicated in the proof, and the new partition function
\be
\tilde{Z}_{RR}(\tau,\bar{\tau}) = Z_{RR}(\tau,\bar{\tau}) - \mathcal{M}_{\alpha \alpha} c_\alpha^2
\ee
remains $\text{SL}(2,\mathbb{Z})$ invariant. Applying the strategy of the proof to $\tilde{Z}_{RR}(\tau,\bar{\tau})$ entails that it ought be zero in order to avoid contradiction. Therefore, the original partition function becomes a constant, and the only non-vanishing RR character is $f_\alpha = c_\alpha$. This indicates a perfect cancellation between bosonic and fermionic excited states, which is a strong evidence for the presence of supersymmetry.

\paragraph{Further Remark} We conclude with a remark on the the Rademacher expansion, 
which is expected to shed new light on our proposal 
that the bound $h^R\geq c/24$ implies the presence of supersymmetry. 
Let us begin with a brief review on the generalization of the Rademacher expansion to 
a vector-valued modular form of weight $w$. See \cite{Dijkgraaf:2000fq} for the detailed discussion.

When each component of the weight $w$ vector-valued modular form $f_i(q)$
can be expanded in powers of $q$ as \eqref{df}, the Fourier 
coefficient $F_i(a)$ has the contour integral representation below,  
\begin{align}
\label{contourintegral}
F_i(a) = \int_{\cal C} d\tau~ e^{-2\pi i \tau (a+m_i)} f_i(\tau), \quad 
\end{align}
where the contour ${\cal C}$ is given by a straight line from $\tau = i$ to $\tau = i+1$. 
To evaluate the integral \eqref{contourintegral}, we deform $\cal C$ to the Rademacher contour ${\cal C}_R(M)$ for a given integer $M$. 
The Rademacher contour is defined in terms of the Ford circle $C(s,d)$, a circle of radius $\frac{1}{2s^2}$ and tangent to the $x$-axis at a Farey number $d/s \in \mathcal{F}_M$. Here $d$ and $s$ are coprime integers such that $d/s$ is a irreducible fraction, and  
the Farey numbers ${\cal F}_M$ include the fractions between 0 and 1 with denominator less than $M$. The Rademacher contour can then be described as follows, 
\begin{align}
{\cal C}_R(M) = \bigcup_{\frac{d}{s} \in \mathcal{F}_M} {\cal C}_{s,d}(M),
\end{align}
where ${\cal C}_{s,d}(M)$ is the arc between the intersection points $\alpha_-(s,d)$ and $\alpha_+(s,d)$ of the Ford circle $C(s,d)$ with the neighboring Ford circles. To illustrate, 
we show the Rademacher contour with $M=4$ in figure \ref{Rcontour}.
Note that ${\cal C}_R (\infty)$ covers the entire arc of the Ford circles except tangential points on the real line. 
\begin{center}
\begin{figure}[!t]
\centering
\begin{tikzpicture}[scale=0.8]
\draw[->] (0.075,-1.0) -- (0.075,9.0);  
\draw[->] (-1.0,0) -- (9.0,0);   

\node [below right] at (2,0) {$\frac{1}{4}$};
\node [below right] at (2.667,0) {$\frac{1}{3}$};
\node [below right] at (4.0,0) {$\frac{1}{2}$};
\node [below right] at (5.333,0) {$\frac{2}{3}$};
\node [below right] at (6,0) {$\frac{3}{4}$};

\node [above left] at (2.30,0.47) {$\alpha_-(1,4)$};
\node [above right] at (5.90,0.47) {$\alpha_+(3,4)$};
 
\node at (2,0) {$\times$}; 
\node at (2.667,0) {$\times$}; 
\node at (4,0) {$\times$}; 
\node at (5.333,0) {$\times$}; 
\node at (6.0,0) {$\times$}; 


\draw[thick,blue,xshift=2pt,
decoration={ markings,
      mark=at position 0.2 with {\arrow{latex}}, 
      mark=at position 0.4 with {\arrow{latex}},
      mark=at position 0.6 with {\arrow{latex}}, 
      mark=at position 0.8 with {\arrow{latex}}},
      postaction={decorate}]
 (0,8.0) -- (8.0,8.0);
\draw[thick,gray,xshift=2pt,
decoration={markings}, postaction={decorate}]
  (0,0) -- (0,8.0);
\draw[thick,gray,xshift=2pt,
decoration={markings}, postaction={decorate}]
  (8.0,0) -- (8.0,8.0);
\draw[thick,gray,xshift=2pt,
decoration={ markings}, 
      postaction={decorate}]
 (0,4.0) [partial ellipse=90:-90:4.0cm and 4cm];
\draw[thick,gray,xshift=2pt,
decoration={ markings}, 
      postaction={decorate}]
 (8.0,4.0) [partial ellipse=270:90:4.0cm and 4cm]; 
 \draw[thick,gray,xshift=2pt,
decoration={ markings}, 
      postaction={decorate}]
 (4.0,1.0) [partial ellipse=360:0:1.0cm and 1.0cm];
\draw[thick,gray,xshift=2pt,
decoration={ markings}, 
      postaction={decorate}]
  (2.667,0.444) [partial ellipse=360:0:0.444cm and 0.444cm];
\draw[thick,gray,xshift=2pt,
decoration={ markings}, 
      postaction={decorate}]
  (5.333,0.444) [partial ellipse=360:0:0.444cm and 0.444cm];
\draw[thick,gray,xshift=2pt,
decoration={markings},  postaction={decorate}]
  (2.0,0.25) [partial ellipse=0:360:0.25cm and 0.25cm];
\draw[thick,gray,xshift=2pt,
decoration={markings}, postaction={decorate}]
  (6.0,0.25) [partial ellipse=0:360:0.25cm and 0.25cm];

\draw[thick,red,xshift=2pt,
decoration={ markings,
      mark=at position 0.2 with {\arrow{latex}}, 
      mark=at position 0.4 with {\arrow{latex}},
      mark=at position 0.6 with {\arrow{latex}}, 
      mark=at position 0.83 with {\arrow{latex}}}, 
      postaction={decorate}]
 (0,4.0) [partial ellipse=90:-63:4.0cm and 4cm]; 
\draw[thick,red,xshift=2pt,
decoration={ markings,
      mark=at position 0.18 with {\arrow{latex}}, 
      mark=at position 0.4 with {\arrow{latex}},
      mark=at position 0.6 with {\arrow{latex}}, 
      mark=at position 0.8 with {\arrow{latex}}}, 
      postaction={decorate}]
 (8.0,4.0) [partial ellipse=240:90:4.0cm and 4cm];
\draw[thick,red,xshift=2pt,
decoration={ markings,
      mark=at position 0.55 with {\arrow{latex}}, 
      mark=at position 0.77 with {\arrow{latex}}, 
      mark=at position 0.99 with {\arrow{latex}}}, 
      postaction={decorate}]
 (4.0,1.0) [partial ellipse=200:-20:1.0cm and 1.0cm];
\draw[thick,red,xshift=2pt,
decoration={ markings, 
      mark=at position 0.8 with {\arrow{latex}}}, 
      postaction={decorate}]
  (2.667,0.444) [partial ellipse=200:20:0.444cm and 0.444cm];
\draw[thick,red,xshift=2pt,
decoration={ markings, 
      mark=at position 0.8 with {\arrow{latex}}}, 
      postaction={decorate}]
  (5.333,0.444) [partial ellipse=160:-20:0.444cm and 0.444cm];
\draw[thick,red,xshift=2pt,
decoration={markings},  postaction={decorate}]
  (2.0,0.25) [partial ellipse=0:150:0.25cm and 0.25cm];
\draw[thick,red,xshift=2pt,
decoration={markings}, postaction={decorate}]
  (6.0,0.25) [partial ellipse=40:150:0.25cm and 0.25cm];

\node at (1.92,0.444) {$\bullet$}; 
\node at (6.18,0.444) {$\bullet$}; 
\end{tikzpicture}
\caption{\label{Rcontour} We deform the original contour ${\cal C}$ (blue) to the Rademacher contour ${\cal C}_R(M)$ (red) in $\tau$ plane. As an example, we present the Rademacher contour with $M=4$. The number of the Ford circles increases for large $M$, therefore the Rademacher contour covers the whole arc of the Ford circles except the tangent points at the real axis. $\alpha_+(s,d)$ and $\alpha_-(s,d)$ are intersection points of the Ford circle $C(s,d)$ with its neighboring Ford circles. }
\end{figure}
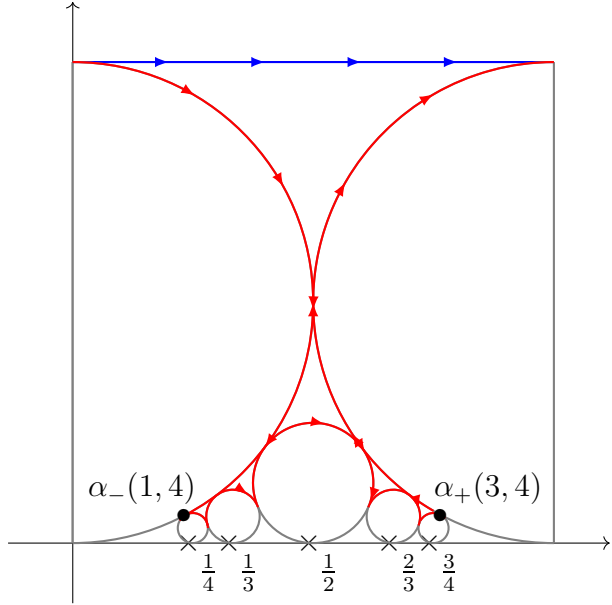
\end{center}

\begin{spacing}{-3.0}
\end{spacing} 

Evaluating the contour integral over ${\cal C}_R(\infty)$, the
Fourier coefficient can be expressed as 
\begin{align}
\label{Rademacher}
\begin{split}
F_{i}(a)&=2\pi \sum_{s=1}^{\infty} \Bigg[ \sum_{i=0}^{N-1} s^{w-2} K\ell(a,i,b,j;s) \sum_{b+m_j <0} F_j(b) \big(2\pi |b+m_j|\big)^{1-w}  \\
& \times \left(\frac{2\pi}{s} \sqrt{(a+m_i)|b+m_j|}\right)^{w-1} I_{1-w}\left( \frac{4\pi}{s} \sqrt{(a+m_i)|b+m_j|}\right) \Bigg] + {\tilde F}_i(a), 
\end{split}
\end{align}
where $K\ell(a,i,b,j;s)$ denotes the Kloosterman sum and $I_\alpha(x)$ is the modified Bessel function of the first kind. It was shown in \cite{Dijkgraaf:2000fq} that 
${\tilde F}_i(a)$ vanishes when the Fourier coefficients $F_j(b)$ with $(b+m_j) \geq 0$ 
are all positive for any $j$ and $w\geq 0$. However, since the Fourier coefficients of the RR characters 
are not necessarily positive, more elaboration is required to show whether ${\tilde F}_i(a) =0$ 
in our case.  If this is the case, the Rademacher expansion can provide 
an alternative argument that the RR conformal characters should be trivial 
when the bound \eqref{unitary bound} is obeyed. 
We leave it to the future work.

\section*{Acknowledgment} 
We wish to thank Kimyeong Lee for helpful discussions. The work of J.B. is supported in part by the European Research Council(ERC) under the European Union's Horizon 2020 research and innovation programme (Grant No. 787185). Z.D. and S.L. are supported in part by KIAS Individual Grant PG076902 and PG056502, respectively.

\bibliographystyle{IEEEtran}
\bibliography{main}

\end{document}